\documentclass[twocolumn,showpacs,preprintnumbers,amsmath,amssymb,nofootinbib,floatfix]{revtex4-1}
\usepackage{graphicx}% Include figure files
\usepackage{dcolumn}% Align table columns on decimal point
\usepackage{bm}% bold math

\newcommand{\be}{\begin{equation}}
\newcommand{\ee}{\end{equation}}
\newcommand{\ba}{\begin{eqnarray}}
\newcommand{\ea}{\end{eqnarray}}
\newcommand{\bi}{\begin{itemize}}
\newcommand{\ei}{\end{itemize}}
\newcommand{\tr}{{\rm Tr\,}}

\newcommand{\eq}{Eq.~}

\newcommand{\fig}{Fig.~}

\newcommand{\la}{\label}

\newcommand{\txts}{\textstyle}

\newcommand{\bp}{\boldsymbol{p}}

\newcommand{\bx}{\boldsymbol{x}}
\newcommand{\bz}{\boldsymbol{z}}

\newcommand{\Fgen}{K}
\begin{document}

\preprint{CERN-PH-TH/2010-196, MKPH-T-10-31}

\title{Thermal momentum distribution from path integrals with shifted boundary conditions}

\author{Leonardo Giusti$^{a,b}$, Harvey B. Meyer$^c$}

\affiliation{\vspace{0.3cm}
\vspace{0.1cm} $^a$ Physics Department, CERN, CH-1211 Geneva 23, Switzerland\\
$^b$ Dipartimento di Fisica, Universit\'a di Milano Bicocca, Piazza della Scienza 3, 
I-20126 Milano, Italy\\
$^c$ Institut f\"ur Kernphysik, Johannes Gutenberg-Universit\"at Mainz,
D-55099 Mainz, Germany}

\date{\vspace{0.2cm} \today}

\begin{abstract}
For a thermal field theory formulated in the grand canonical ensemble,
the distribution of the total momentum is an observable characterizing
the thermal state. We show that its cumulants are related to thermodynamic
potentials. In a relativistic system for instance, the thermal
variance of the total momentum is a direct measure of the enthalpy.
We relate the generating function of the cumulants to the ratio of (a)
a partition function expressed as a Matsubara path integral with
shifted boundary conditions in the compact direction, and (b) the
ordinary partition function. In this form the generating function is
well suited for Monte-Carlo evaluation, and the cumulants can be
extracted straightforwardly. We test the method in the SU(3)
Yang--Mills theory and obtain the entropy density at three different
temperatures.
\end{abstract}

% PACS, the Physics and Astronomy
                             % Classification Scheme.
%\keywords{Suggested keywords}%Use showkeys class option if keyword
                              %display desired
\maketitle

\section{Introduction\label{sec:intro}}
\noindent 
Thermal field theory is a theoretical tool of central importance in
condensed matter physics, plasma physics, nuclear physics and
cosmology~\cite{Kapusta:2006pm,LeBellac}.  Obtaining first-principles
predictions from a thermal field theory is often challenging, since it
describes an infinite number of degrees of freedom subject to both
quantum and thermal fluctuations.  New theoretical concepts and more
efficient computational techniques are still needed in many contexts,
particularly when weak-coupling methods are inapplicable.

In this work we exploit the global symmetries of a thermal theory to
define the contributions to the partition function of states with
given quantum numbers, and show that they are well suited for \emph{ab
  initio} Monte-Carlo computations.  Our main results, which are based
on spatial translation invariance, concern the relative contribution
to the partition function of states with total momentum $\bp$.  These
contributions form the probability distribution of $\bp$, whose
cumulants can be related in a simple manner to the equation of state
by using Lorentzian or Galilean invariance.  For instance, in Quantum
Chromodynamics (QCD) at zero baryon chemical potential
$\mu$, the variance of the total momentum measures the
entropy of the system.  As we will see shortly, the generating
function $\Fgen$ of the cumulants of the momentum distribution can be
expressed as a ratio of two partition functions. The latter
are represented by Euclidean path integrals with different boundary
conditions for the fields in the time direction, and
their ratio can therefore be computed by standard Monte Carlo
techniques.

While the following theoretical discussion extends to a wide class of
theories, we illustrate our ideas numerically in the $SU(3)$
Yang--Mills theory, where we determine the entropy density of the
system at three different temperatures.  Crucially, the method can be
applied to QCD, since the hermiticity properties of finite-difference
operators are not affected by our `shifted' boundary conditions, and
hence the actions to be used in the simulations are always real at
$\mu=0$.  As an example of an application to a
non-relativistic system, it can also be employed in the study of
neutron matter in the Euclidean approach of~\cite{Lee:2004qd}.

There are already several established methods to compute the
thermodynamic properties of gauge
theories~\cite{Engels:1981qx,Engels:1990vr,Endrodi:2007tq,Meyer:2009tq}.
They require either a vacuum subtraction or a renormalization constant
to be determined, facts which make it difficult to apply them at
arbitrarily high and low temperatures. 
Our method avoids these problems, and has the further advantage 
that a Symanzik improvement of the action (i.e. a
suppression of the discretization
errors~\cite{Symanzik:1983dc,Luscher:1984xn}) automatically leads to a
corresponding improvement in the thermodynamic
quantities. Computationally, the method is rather expensive, since it
consists in calculating a ratio of partition functions. However our
experience shows that, even with a simple algorithm, physics results
can be obtained with commonly available computing resources.

\section{Momentum distribution \label{sec:theory}}
\noindent
Recent progress in lattice field theory has made it possible to define and
compute by Monte Carlo simulations the relative contribution to the
partition function due to states carrying a given set of quantum numbers
associated with exact symmetries of a
theory~\cite{DellaMorte:2008jd,DellaMorte:2010}.  Here we apply these
techniques to a finite temperature and density system in the grand
canonical ensemble (or the canonical ensemble if there is no conserved
particle number).  The relative contribution to the partition function
of the states with momentum $\bp$ is given by
\be
\frac{R(\beta,\mu,{\bp})}{L^3} = \langle \hat{\rm P}^{({\bp})}\rangle = 
\frac{{\rm Tr}\{e^{-\beta (\hat H-\mu\hat N)} 
\hat{\rm P}^{({\bp})}\}}{{\rm Tr}\{e^{-\beta (\hat H-\mu\hat N)}\}}
\ee
where the trace is over all the states of the Hilbert space, $\hat{\rm
  P}^{({\bp})}$ is the projector onto those states with total momentum
$\bp$, $\beta=1/T$ is the inverse temperature, $\hat H$ the Hamiltonian,
$\hat N$ the particle number and $L$ the linear size of the system.
The generating function $\Fgen$ 
associated with the momentum distribution is defined as 
\be\label{eq:freeE}
e^{-\Fgen(\beta,\mu,{\bz})} = 
\frac{1}{L^3} \sum_{\bp} e^{i{\bp}\cdot{\bz}} \, R(\beta,\mu,{\bp})\;.
\ee
The connected cumulants are obtained from it as follows,
\be
\frac{K_{\{2 n_1, 2 n_2, 2 n_3\}}}{(-1)^{n_1+n_2+n_3+1}} = 
{\txts\frac{\partial^{2n_1}}
{\partial \bz_1^{2n_1}} \frac{\partial^{2n_2}}{\partial \bz_2^{2n_2}}
\frac{\partial^{2n_3}}{\partial \bz_3^{2n_3}}}
\frac{\Fgen({\bz})}{L^3}\Big|_{\bz=0}\; ,
\ee
where they have been normalized so as to have a finite limit when $L\rightarrow\infty$
and the $(\beta,\mu)$ dependence has been suppressed.
Finally, we can write 
\be\label{eq:bella}
R(\beta,\mu,{\bp}) = \int d^3{\bz}\, e^{-i{\bp}\cdot{\bz}}\, \frac{Z(\beta,\mu,\bz)}{Z(\beta,\mu)},
\ee
where $Z(\beta,\mu,\bz)=\tr\{e^{-\beta(\hat H-\mu\hat N)+i\hat\bp\bz}\}$
is a partition function in which states of momentum $\bp$ are weighted
by a phase $e^{i\bp\cdot\bz}$.
The shifted partition function $Z(\beta,\mu,\bz)$ 
can be expressed as a path integral in Euclidean time by adopting 
the `shifted' boundary conditions
\be
\phi(\beta,{\bx})=\pm\phi(0,{\bx}+{\bz})\; ~~
\la{eq:bc}
\ee
with the $+(-)$ sign for bosonic (fermionic) fields respectively.
From Eqs.~(\ref{eq:freeE}) and (\ref{eq:bella}),
the generating function can be written
as the ratio of two partition functions,
\be
e^{-\Fgen(\beta,\mu,{\bz})} = \frac{Z(\beta,\mu,\bz)}{Z(\beta,\mu)}\;,
\la{eq:exp-F}
\ee 
i.e.~two path integrals with the same action 
but different boundary conditions. 
In a renormalizeable theory it therefore has a finite and universal continuum limit.

We remark that in the large volume regime, the momentum distribution
can be expressed at each value of $\bp$ as a saddle point expansion.
Its leading term in $1/L^3$, for a system with a finite correlation
length, is a Gaussian with a width equal to the second cumulant. Near
a second-order phase transition on the other hand, 
the relative size of the fourth and second cumulants
can serve to characterize a universality class.
\section{Connection to thermodynamics}
In thermal field theories, there is a connection between the
cumulants of the momentum distribution and thermodynamic functions.
In relativistic theories, the reason is that the momentum density
$\pi_i\equiv T_{0i}$ can be chosen to coincide with the energy flux.
In non-relativistic theories, the particles are the only carriers of
momentum, and therefore the momentum density (divided by the mass $m$ of
the particles) coincides with the particle number
flux~\cite{Landau-hydro}. The relation
\ba
&&{\txts\int} d^3\bx \,e^{ik_3x_3}\,
 \langle  \pi_{3}(x)\, \pi_{3}(0)\rangle 
\la{eq:WI} \\
&&= \left\{
\begin{array}{l@{~~}l}
\int d^3\bx \,e^{ik_3x_3}\,\langle T_{00}(x)\, T_{33}(0) \rangle_c & (\textrm{rel.}) 
\\
m\int d^3\bx \,e^{ik_3x_3}\, \langle n(x)\, T_{33}(0) \rangle_c  & (\textrm{non-rel.})
\end{array}\right.
\nonumber
\ea
then follows from the Ward identities (WIs) associated with (a) the
conservation of momentum and (b) the conservation of energy (particle
number) respectively in the relativistic (non-relativistic) case.
\eq(\ref{eq:WI}) is valid for any $x_0\neq0$ and non-vanishing momentum $k_3$ 
in a periodic box of size $L$,
and we now take the limit $L\to\infty$, with $k_3L$ fixed.
In a fluid at rest the pressure $p(T)$ is given by the thermal average
of the stress tensor $\langle  \hat T_{ij} \rangle = \delta_{ij} p(T)$,
and the relations $\frac{\partial p}{\partial\mu}(T,\mu)=n$,
$\frac{\partial p}{\partial T}(T,\mu)=s$, $Ts+\mu n = e+p$ hold 
in the thermodynamic limit. We thus obtain 
\be
K_{2,0,0}(\beta,\mu) =\left\{
\begin{array}{l@{~~~~}l}
T(e+p)
 & (\textrm{rel.}) 
\\
Tm n
  & (\textrm{non-rel.}).
\end{array}\right.
\la{eq:master}
\ee
Alternatively, \eq(\ref{eq:master}) can be derived directly at $k_3=0$
from the same family of WIs. In this way an expression for the finite-volume correction
is obtained~\cite{GM:2010}.
% using the linear response formulae and adopting
% hydrodynamics as a low-momentum description~\cite{Kadanoff-Martin,Teaney:2006nc,Minami:2009hn}.

Relativistic theories at $\mu=0$ constitute an important special case
that we investigate in more detail, since it is relevant both 
to heavy-ion collisions~\cite{Teaney:2009qa} 
and to the physics of the early universe~\cite{McGuigan:2008pz,Sassi:2009tr}.
First, the relation $e+p=Ts$ implies that
the thermal variance of the momentum is a direct measure of 
the entropy of the system. 
Secondly, by establishing a recursion relation among the cumulants
based on the Ward identities, it is possible to show that the 
fourth order cumulants are related to the specific heat of the 
system~\cite{GM:2010},
\be
c_v = \frac{K_{4,0,0}}{3 T^4} - \frac{3\, K_{2,0,0}}{T^2}
=\frac{K_{2,2,0}}{T^4} - \frac{3\, K_{2,0,0}}{T^2}.
\la{eq:cv}
\ee
Combining second and fourth cumulants, one can thus obtain the 
speed of sound $c_s^2=\frac{s}{c_v}$.
One may prove~\cite{GM:2010} that in conformal field theories, 
the generating function is completely determined by its first non-zero cumulant,
\be
\frac{\Fgen_{\mbox{\tiny CFT}}(\beta,{\bz})}{L^3} = \frac{s_{\mbox{\tiny CFT}}(T)}{4}
\left(1-\frac{1}{(1+T^2 {\bz}^2)^2} \right).
\la{eq:sCFT}
\ee
Thirdly, it can be shown~\cite{GM:2010} that finite volume effects in $K_{2,0,0}$ are
exponentially small in $m(T)\cdot L$, where $m(T)$ is the lightest
screening mass in the theory, and the prefactor is explicitly
known~\cite{Meyer:2009kn}.  In summary, the cumulants can be used to
calculate thermodynamic properties.

\begin{table}[t!]
\begin{center}
\begin{tabular}{llcccccc}
\hline\\[-0.175cm]
Lat        &$6/g^2_0$&$\beta/a$&$L/a$& $\frac{1}{n}(\frac{L}{a})^3$&$r_0/a$&$K(\beta,{\bz},a)$ & 
$\frac{2 \Fgen(\beta, \bz, a)}{{|{\bz}|}^2 T^5 L^3}$ % $s_{\rm eff}(T)/T^3$ 
\\[0.125cm]
\hline\\[-0.175cm]
${\rm A}_1$   &$5.9$  & $4$&$12$& 2 & $4.48(5)$&$17.20(11)$&$5.10(3)$ \\[0.125cm]
${\rm A}_{1a}$&$5.9$  & $4$&$16$& 2 & $4.48(5)$&$40.71(15)$&$5.089(19)$ \\[0.125cm] %
${\rm A}_2$   &$6.024$& $5$&$16$& 2 & $5.58(6)$&$13.05(10)$&$4.98(4)$ \\[0.125cm]
${\rm A}_3$   &$6.137$& $6$&$18$& 3 & $6.69(7)$&$7.32(8)$  &$4.88(6)$ \\[0.125cm]
${\rm A}_4$   &$6.337$& $8$&$24$& 4 & $8.96(9)$&$4.32(16)$ &$5.12(19)$\\[0.125cm]
${\rm A}_5$   &$6.507$&$10$&$30$& 5 & $11.29(11)$&$2.62(17)$ &$4.9(3)$\\[0.20cm] % 
\hline\\[-0.175cm]
${\rm B}_1$   &$6.572$& $4$&$12$& 2 & $12.28(12)$&$22.22(11)$&$6.58(3)$ \\[0.125cm]
${\rm B}_{1a}$&$6.572$& $4$&$16$& 2 & $12.28(12)$&$53.47(16)$&$6.684(20)$ \\[0.125cm] %
${\rm B}_2$   &$6.747$& $5$&$16$& 2 & $15.34(15)$&$17.11(15)$&$6.53(6)$ \\[0.125cm]
${\rm B}_3$   &$6.883$& $6$&$18$& 3 & $18.14(18)$&$9.61(9)$  &$6.40(6) $\\[0.125cm]
${\rm B}_4$   &$7.135$& $8$&$24$& 4 & $24.5(3)$&$5.42(17)$ &$6.42(20)$\\[0.125cm]
${\rm B}_5$   &$7.325$&$10$&$30$& 5 & $30.7(4)$&$3.32(18)$ &$6.1(3)$\\[0.20cm]%
\hline\\[-0.175cm]
${\rm C}_1$   &$7.234$& $4$&$16$& 4 & $27.6(3)$&$57.44(25)$&$7.18(3)$ \\[0.125cm]
${\rm C}_2$   &$7.426$& $5$&$20$& 5 & $34.5(4)$&$36.5(4)$&$7.13(8)$ \\[0.125cm]
${\rm C}_3$   &$7.584$& $6$&$24$& 4 & $41.4(5)$&$24.7(4)$&$6.94(12)$ \\[0.125cm]
\hline
\end{tabular}
\caption{
\label{tab:lattices} Lattice parameters and numerical results with $\bz=(2,0,0)$. 
The quantity in the last column is the estimator for the entropy density, see
  \eq(\ref{eq:sT3}).}
\end{center}
\end{table}

\section{Numerical implementation and results\label{sec:algo}}

\noindent There are many potential applications of formulae
(\ref{eq:master},\ref{eq:cv}).  As an illustration we consider the
$SU(3)$ lattice Yang--Mills theory defined in a cubic box of volume
$L^3$ with periodic boundary conditions, at temperature $\beta=1/T$
and lattice spacing $a$.  Our goal is to show that the entropy density
can be computed in the thermodynamic and continuum limit.

Even though the lattice breaks continuous translation invariance,
discrete translations remain as exact symmetries of the lattice
action. They suffice to define $\Fgen$ at finite lattice spacing
precisely as in section (\ref{sec:theory}).  The function
$\Fgen(\beta,\bz,a)$ defined on the lattice differs by O($a^2$)
effects from its continuum counterpart.

The canonical partition function is given as usual by 
\be\label{eq:stPF}
Z(\beta) = {\rm Tr}\{e^{-\beta\hat H}\} = \int D[U]\, e^{-S(U)}\;,
\ee
where we suppress the dependence of $Z$ on $a$ and $D[U]$ is the Haar
measure over the gauge links $U_\mu(x_0,{\bx})$, for $\mu=0,1,2,3$ and
for each lattice point $(x_0,{\bx})$. For definiteness we choose the
action $S(U)$ to be the standard Wilson plaquette action
\cite{Wilson:1974sk}, whose bare coupling we denote by $g_0^2$ (for
unexplained notation see Ref.~\cite{DellaMorte:2008jd}).

The most straightforward way to compute $Z(\beta,\bz)/Z(\beta)$ as
given in \eq(\ref{eq:exp-F}) is to define a set of $(n+1)$ systems
with partition functions ${\cal Z}(\beta,r_i)$, ($r_i=i/n$, $i=0,1,\dots,n$)
and corresponding actions $\overline S(U,r_i)= r_i S(U) + (1-r_i) S(U^z)$,
where $U^z$ is the field obeying shifted boundary conditions, see \eq(\ref{eq:bc}).
We can then write the identity
\be\label{eq:prod}
\frac{Z(\beta,\bz)}{Z(\beta)} = \prod_{i=0}^{n-1}
\frac{{\cal Z}(\beta,r_i)}{{\cal Z}(\beta,r_{i+1})}\; .
\ee
If one defines the `reweighting' observable as 
\be
O(U,r_{i+1}) = 
e^{{\overline S}(U,r_{i+1})-{\overline S}(U,r_i)}\; ,
\ee
then the ratio ${\cal Z}(\beta,r_i)/{\cal Z}(\beta,r_{i+1})$ in 
Eq.~(\ref{eq:prod}) can be computed as its expectation 
value on the ensemble of gauge 
configurations generated with the action 
${\overline S}(U,r_{i+1})$~\cite{DellaMorte:2008jd}. The generating functional 
can thus be written as 
\be
\Fgen(\beta,\bz,a) = -\sum_{i=0}^{n-1} 
\ln{\left\{\frac{{\cal Z}(\beta,r_i)}{{\cal Z}(\beta,r_{i+1})}\right\}}\; . 
\ee
The continuum limit value of the second cumulant
is obtained from a discrete lattice by a limiting procedure,
\be
\frac{s(T)}{T^3} = \frac{K_{2,0,0}(\beta)}{T^5} = 
\lim_{a\rightarrow 0} 
\frac{2 \Fgen(\beta, \bz, a)}{{|{\bz}|}^2 T^5 L^3}\; ,
\la{eq:sT3}
\ee
where $\bz=(n_z a,0,0)$, the integer $n_z$ being kept fixed 
when $a \rightarrow 0$. The continuum value of the entropy is 
thus approached with O($a^2$) corrections.

If the set of interpolating systems is chosen so that the error of
each contribution in the sum is comparable and $n=(L/a)^3$, then the
cost of the simulations for a given relative precision on
$\Fgen(\beta,{\bz},a)$, at fixed value of ${\bz}$, scales
approximatively as $a^{-7}$, while it remains roughly constant as a
function of $L$. Each derivative of $\Fgen$ at the origin requires an
extra factor $\propto a^{-2}$ in statistics and therefore in the
overall cost. Such a cost figure does not take into account
autocorrelation times, and it depends heavily on the particular
algorithm that we have implemented. For instance, a more refined
interpolation path between the initial and final systems may lead to a
significant speedup.

\begin{figure}[t!]
\begin{center}
\includegraphics*[angle=-90,width=7.8cm]{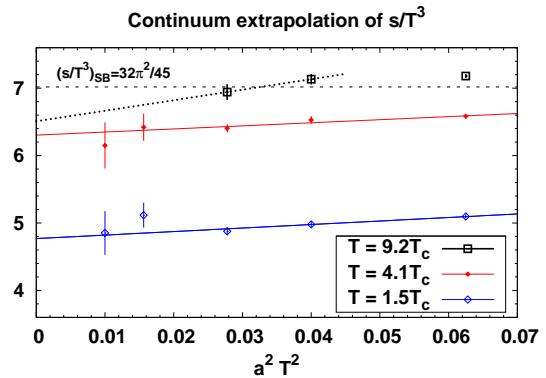}
\vspace{-0.4cm}
\caption{\label{fig:conti} Scaling behavior of $s/T^3$, 
see \eq(\ref{eq:sT3}). The Stefan-Boltzmann value reached 
in the high-$T$ limit is also displayed.}
\end{center}
\end{figure}

We have calculated the entropy at three temperatures, 1.5, 4.1 and
9.2$T_c$, see Table \ref{tab:lattices} for the numerical results.
The update algorithm used is the standard combination of heatbath and
overrelaxation
sweeps~\cite{Fabricius:1984wp,Cabibbo:1982zn,Kennedy:1985nu}. The only
changes over the standard algorithm reside (a) in the computation of
the `staples' that determine the contribution to the action of a given
link variable $U_\mu(x)$ and (b) in the more frequent updating of the
two time-slices on which the observable has its support.

The bare coupling $g_0^2$ was tuned using the data
of~\cite{Necco:2001xg} in order to match lattices of different
$\beta/a$ to the same temperature.  Motivated by a study of the free
case, we chose $n_z=2$. The scaling behavior of the entropy density is
displayed in \fig\ref{fig:conti}. We observe that the cutoff effects
are quite mild. Taking into account the systematic uncertainty in
performing the continuum extrapolation, our results at the
two lower temperatures are compatible with the published
results~\cite{Boyd:1996bx,Namekawa:2001ih}.
% At $1.5T_c$ we find $4.77(8)$, while \cite{Namekawa:2001ih} 
% obtained about 4.92(20) and \cite{Boyd:1996bx} 5.07(20).
The results at $9.2T_c$ are new.

\section{Final Remarks}
In this Letter we have introduced the generating function of
cumulants of the total-momentum distribution 
and a new way of computing it.
As an application, we have shown that the second cumulant is a measure of
the enthalpy (or particle density in the non-relativistic case, see
\eq(\ref{eq:master})), and calculated the enthalpy density
of gluons at three different temperatures.  
The ideas presented in this paper have further
interesting applications. 

One application is the determination of the partial pressures of
different symmetry sectors (labeled by continuous as well as discrete
quantum numbers) in the confined phase of QCD. This will provide a far
more stringent test of the hadron resonance gas model than has been
possible so far. The latter model postulates that the QCD pressure is
due to the sum of the partial pressures of all zero-temperature
resonances of width $\Gamma\lesssim T$, and is an important ingredient
in the phenomenology of heavy-ion collisions (see for
instance~\cite{Andronic:2008gu}).  Beyond the pressure, any observable
can be calculated in the restricted ensemble where certain quantum
numbers assume prescribed values.

Our method has a `kinematic' character, and the lattice action is the
only ingredient in the calculation. It is conceivable that the
boundary conditions adopted in this paper can be used to formulate
Symanzik improvement conditions~\cite{Symanzik:1983dc,Luscher:1984xn},
or to compute the constants needed to define an energy-momentum tensor
which satisfies the Ward identities in the continuum
limit~\cite{Caracciolo:1989pt}.

We stress that the generating function $\Fgen(\beta,\mu,\bz)$ is of
intrinsic interest, beyond giving access to basic thermodynamic
quantities. It can be used to assess how nearly scale-invariant the
system is at a given temperature, see \eq(\ref{eq:sCFT}).  In QCD at
finite baryon density and in the low-temperature limit, the generating
function $\Fgen$ is an order parameter for the spontaneous breaking of
translation invariance. Finally, it may be of interest in analytic
treatments, where it is more convenient to absorb the shift $\bz$ into
the action~\cite{Kapusta:1981ay}. The propagators are then modified,
the parameter $i\bz/\beta$ playing the role of an external velocity
parameter.

\section*{Acknowledgments\label{sec:ack}}
We thank Michele Della Morte and Martin L\"uscher for useful
discussions.  The simulations were performed on PC clusters at CERN,
at the CILEA, at the CSCS and on the Wilson cluster in Mainz.  We
thankfully acknowledge the computer resources and technical support
provided by all these institutions and their technical staff.

\bibliography{../lattice}% Produces the bibliography via BibTeX.
\end{document}